\documentclass[prb,twocolumn,superscriptaddress,preprintnumbers,amsmath,amssymb]{revtex4}
\usepackage{graphicx}
\usepackage[colorlinks=true,citecolor=blue]{hyperref}

\begin{document}
    \title{Single-photon blockade in doubly resonant nanocavities with second-order nonlinearity}
    \author{Arka Majumdar}
   \email{arka.majumdar@berkeley.edu}
    \affiliation{Department of Physics, University of California at Berkeley, CA-94720, United States}
    \author{Dario Gerace}
    \email{dario.gerace@unipv.it}
    \affiliation{Department of Physics, University of Pavia, I-27100 Pavia, Italy}


\begin{abstract}
We propose the use of nanostructured photonic nanocavities made of $\chi^{(2)}$ nonlinear materials as prospective passive devices to generate strongly sub-Poissonian light via single-photon blockade of an input coherent field. The simplest scheme is based on the requirement that the nanocavity be doubly resonant, i.e. possess cavity modes with good spatial overlap at both the fundamental and second-harmonic frequencies. We discuss feasibility of this scheme with state-of-the art nanofabrication technology, and the possibility to use it as a passive single-photon source  on-demand.
\end{abstract}
\maketitle

\section{Introduction}

Several prospective applications in quantum information science employing photonic systems require the ability to efficiently generate indistinguishable single-photon states, possibly on-demand \cite{mete2004,qp_review,focus_iqo}. To this end, significant progress has been made to achieve single-photon emission from non-resonantly driven solid-state quantum emitters \cite{auffeves2010}, such as single quantum dots \cite{sps_review} or impurities in solids \cite{kurtsiefer00prl}, most often coupled to photonic nanostructures such as waveguides \cite{claudon2010,shields2011apl,finley2012prx} or cavities \cite{michler00sci,Pelton2002,Santori2002,englund2005prl,kevin07nat,faraon2011nphot} for increased emission rate and collection efficiency. Alternatively, since non-resonant pumping results in a reduced indistinguishability between the sequentially emitted single photons, the same effect can be obtained by resonant coherent driving of an emitter-cavity coupled system inducing \textit{conventional} single-photon blockade \cite{imamoglu99}: when a coherent light beam enters a nonlinear system whose effective nonlinear response is able to produce a shift of the two-photon resonance that is larger than its linewidth, it generates a stream of single photons at the output of the device. Such an effect has been shown both with atomic \cite{birnbaum05nat} and solid state cavity quantum electrodynamics (cQED) systems based on single quantum dots in photonic resonators \cite{faraon08nphys,volz2012nphot,mete2013nnano}. Several proposals have been recently put forward to achieve conventional photon blockade with enhanced resonant nonlinearities in confined photonic systems \cite{ciuti06prb,chang07np,gerace_josephson,carusotto09epl,arka2013njp}. In addition, a mechanism of \textit{unconventional} photon blockade has also been proposed \cite{savona10prl,bamba,arka2012} to relax the requirements on the system nonlinearity, essentially relying on destructive quantum interference between coupled modes that - under suitable pump/detection conditions - could suppress two-photon emission from the system. However, no conclusive experimental demonstration has been shown confirming such proposals to date.

Open issues towards a fully integrated quantum photonic technology include the scalability on one side, and the possibility to work at telecommunication wavelengths for a direct interfacing with long distance communication networks on the other. In the former case, scalable single-photon sources based on single quantum emitters, e.g. semiconductor quantum dots, might be hindered by difficulties in deterministic positioning, and their degree of inhomogeneity. In the second, the development of efficient single-photon emitters at telecommunication wavelengths is still at its beginning \cite{zinoni,notomi2012scirep}. For these reasons, a promising alternative would be to directly generate single-photon outputs from bulk material nonlinearities, such as the second- or third-order sub-bandgap electronic contributions to the nonlinear polarization of semiconductors or insulators \cite{boyd_book}.
Nonlinear susceptibilities in ordinary semiconductors employed in nanophotonics applications, such as silicon (Si), are too small to produce single-photon nonlinear behavior. Nevertheless, for centrosymmetric materials with $\chi^{(3)}\neq 0$,  it has been recently proposed that the ultimate regime of single-photon blockade might be within reach both for a conventional \cite{ferretti2012} and an unconventional \cite{ferretti2013} scheme, thanks to the unprecedented progress in nanofabrication technology that allows to realize mid-infrared nanocavities with sub-diffraction limited mode volumes, $V_{\mathrm{mode}}$, and ultra-high quality factors, $Q$ (see, e.g., \onlinecite{notomi_review}).

\begin{figure}[t]
\begin{center}
\includegraphics[width=0.48\textwidth]{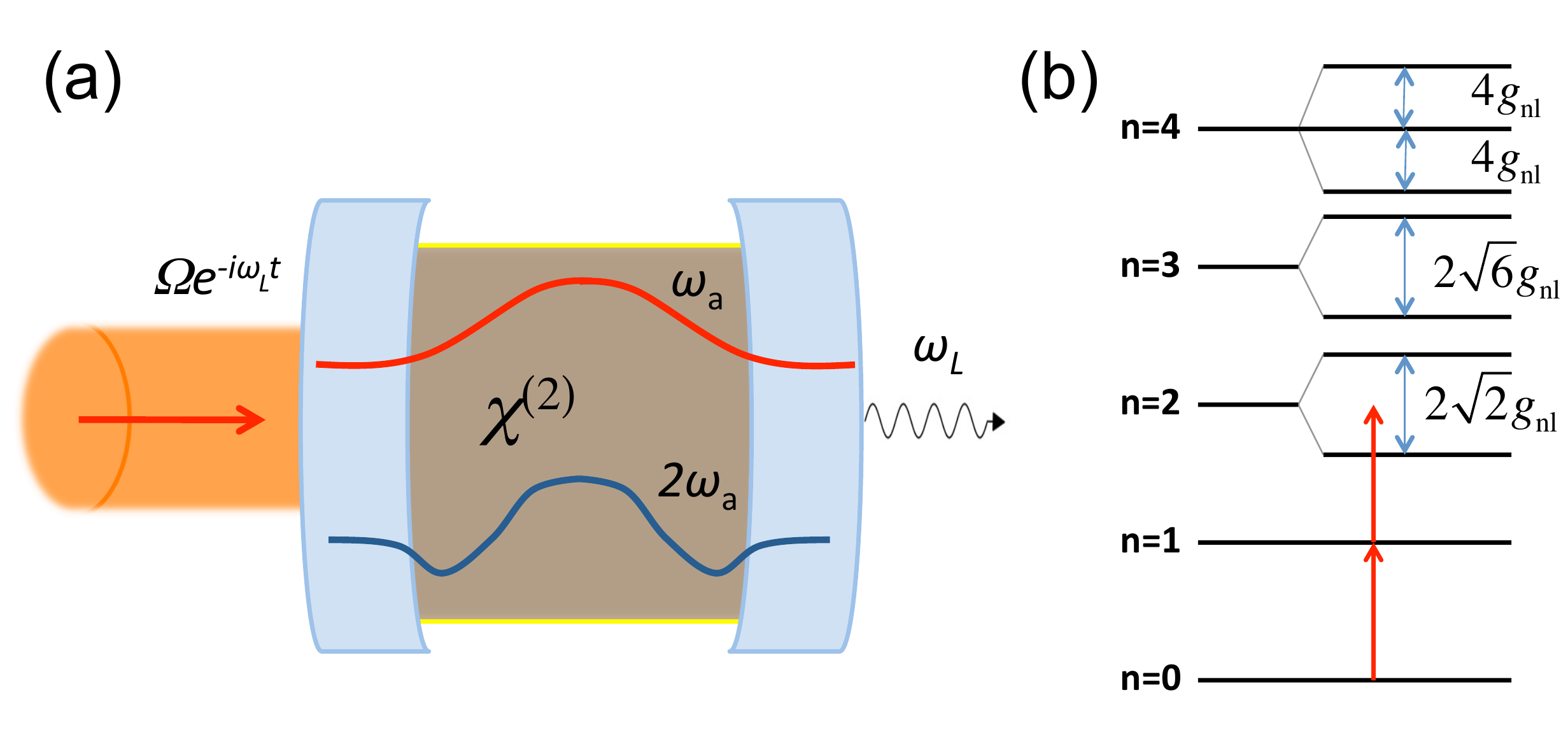}
\caption{(Color online) (a) Scheme of a doubly resonant nanocavity made of a $\chi^{(2)}$
   nonlinear material, which is driven by a coherent field nearly resonant to its fundamental frequency and undergoes
   single-photon blockade at the driving laser frequency. {(b) The schematic energy spectrum of the doubly-resonant nonlinear cavity system (see the Hamiltonian in Eq.~\ref{system_ham}). The arrows show the frequency of the resonantly driving laser.}} \label{fig1}
\end{center}
\end{figure}

Here we extend the previous treatment to second-order nonlinear materials with a susceptibility $\chi^{(2)}\neq 0$, which is much larger than $\chi^{(3)}$ and potentially allows to relax some of the stringent requirements on cavity Q-factors \cite{ferretti2012}. In this case, applications to non-centrosymmetric materials (such as GaP, or GaAs) are implicitly assumed \cite{rivoire2011opex}, although strain- or surface-induced $\chi^{(2)}$ can be enhanced also in typical centrosymmetric materials, such as silicon \cite{matteo_shg_thg}. We apply the quantum master equation for the density matrix of the system, with fundamental and second-harmonic frequencies assumed as high-Q and mode-overlapping resonances. Such doubly resonant systems \cite{liscidini04apl,johnson2007opex}, which might be realized either with two overlapping single-mode cavities or with two different harmonics of the same cavity (see schematic picture in Fig.~\ref{fig1}a), possess an effective photon-photon interaction due to confinement-enhanced bulk $\chi^{(2)}$ nonlinearity. A similar model has already been discussed in the literature with a focus on strong coupling between single photons \cite{carusotto99prb,irvine06prl}. We go beyond previous works by numerically solving the master equation under coherent resonant driving, focussing on the statistics of the emitted photons, and we show that this scheme allows to obtain \textit{conventional} single-photon blockade with state-of-art system parameters. We finally explore the potential use of this passive device as a single-photon source on demand, where the output of the quantum nonlinear source of light is solely determined by the resonance wavelength of the nanocavities, which can be tuned to the telecommunication band by suitable design, and for which the limitations on scalability are sensitively relaxed.

\section{Theory}\label{sec:theory}

We adopt the classical nonlinear optics notation in SI units \cite{boyd_book}.
For a $\chi^{(2)}$ material, the optical response to an applied electromagnetic field is described by
\begin{eqnarray}\label{eq:induction}
D_i(\mathbf{r},t) &=& \varepsilon_{0} \varepsilon_{ij}(\mathbf{r}) E_{j}(\mathbf{r},t)  + \nonumber \\
&+& \varepsilon_{0}[ \chi^{(2)}_{ijk}(\mathbf{r})E_{j}(\mathbf{r},t)E_{k}(\mathbf{r},t)
+O(\chi^{(3)})]  \, ,
\end{eqnarray}
where the indices run over the three spatial directions.
Specifying the nonlinear response to the case of two modes of the electromagnetic field,
and assuming dielectric inhomogeneity but isotropic response for simplicity
(i.e., we will assume scalar linear and nonlinear susceptibilities, $\varepsilon(\mathbf{r})$ and $\chi^{(2)}(\mathbf{r})$),
canonical quantization can be formulated after expressing the field operators as
\begin{eqnarray}\label{fields}
&& \hat{\mathbf{E}}(\mathbf{r},t) =  \nonumber \\
&& \sum_{i=a,b} i \sqrt{\frac{\hbar\omega_i}{2\epsilon_{0}} }
\left[\hat{A}_i \frac{{\vec{\alpha}_i}(\mathbf{r})}{\sqrt{\varepsilon(\mathbf{r})}} e^{-i \omega_i t}-
\hat{A}_i^{\dagger} \frac{{\vec{\alpha}_i}^{\ast}(\mathbf{r})}{\sqrt{\varepsilon(\mathbf{r})}} e^{i \omega_i t}
\right]       \,
\end{eqnarray}
and $\hat{\mathbf{B}}(\mathbf{r})=(-i/ \omega)\nabla\times \hat{\mathbf{E}}(\mathbf{r})$,
where $\hat{A}_a = \hat{a} $ ($\hat{a}^{\dagger}$) and $\hat{A}_b = \hat{b}$ ($\hat{b}^{\dagger}$) define the
destruction (creation) operators of single photon quantum in the modes oscillating at $\omega_a$ and $\omega_b = 2 \omega_a$, respectively.
In this general formalism, Eq. \ref{fields} fully takes into account the spatial dependence of the dielectric environment, such
as in actual nanostructured semiconductor cavities \cite{ferretti2012}.
For each of the two modes, the three dimensional cavity field is normalized according to $\int |\vec{\alpha}_i(\mathbf{r})|^2 \mathrm{d}^3\mathbf{r} =1$
($i=a,b$). Thus, the second-quantized system hamiltonian {up to nonlinear leading orders} is derived from the expression of the time-averaged electromagnetic energy
density,
$\mathcal{H}_{em}=
\frac{1}{2}\int\left[\mathbf{E}(\mathbf{r})\cdot\mathbf{D}(\mathbf{r})
+ \mathbf{H}(\mathbf{r})\cdot\mathbf{B}(\mathbf{r}) \right] \mathrm{d}^3\mathbf{r} \, $
where $\mathbf{H}=\mathbf{B}/ \mu_0$, as the nonlinear expression \cite{carusotto99prb,irvine06prl}
\begin{equation}\label{system_ham}
\hat{H}_s=\hbar\omega_a \hat{a}^{\dagger}\hat{a} + \hbar\omega_b \hat{b}^{\dagger}\hat{b} +
\hbar g_{\mathrm{nl}} [\hat{b}(\hat{a}^\dag)^2+\hat{b}^\dag \hat{a}^2 ] \, ,
\end{equation}
where the nonlinear interaction coefficient can be expressed in terms of the classical electric
field profiles of the two modes
\begin{equation}\label{shift_chi2}
\hbar g_{\mathrm{nl}} = D \varepsilon_0 \left(\frac{\hbar\omega_a }{2\varepsilon_0}\right) \sqrt{ \frac{\hbar\omega_b }{2\varepsilon_0} }
\int \mathrm{d}\mathbf{r} \,\, \frac{ \chi^{(2)}(\mathbf{r})} {[\varepsilon(\mathbf{r})]^{3/2}}
{\alpha}^{2}_a (\mathbf{r}) {\alpha}_b (\mathbf{r})
  \, .
\end{equation}
For simplicity, we are assuming here that the scalar field profiles are effectively given by the relevant components
of the vectorial fields at fundamental and second-harmonic frequencies, as coupled by the non-zero elements of the
$\chi^{(2)}_{ijk}$ tensor. We stress that any quantitative discussion must
be specified to the material under consideration, its crystalline orientation, and the exact spatial distribution of the field components
for the two cavity modes \cite{irvine06prl}. In Eq.~\ref{shift_chi2} this is shortly described by addition of a degeneracy factor ($D$)
accounting for the number of equivalent terms contracted in the second-order nonlinear contribution of Eq.~\ref{eq:induction} (see
Ref.~\onlinecite{sipe87prb} for a detailed analysis). 
{From diagonalization of the Hamiltonian in Eq.~\ref{system_ham}, the schematic level diagram for the first few excitation manifolds can be analytically found as shown in Fig.~\ref{fig1}b. The preferential basis is given by states of the form $|n_a,n_b\rangle$, where $n_a$ and $n_b$ are the number of quanta in the cavity modes $a$ and $b$, respectively. The ground state is $|0,0\rangle$ and the first excited state is $|1,0\rangle$. However, the bare states $|2,0\rangle$ and $|0,1\rangle$ are degenerate and mix to give a splitting of $2\sqrt{2} g_{\mathrm{nl}}$ between the diagonal eigenstates in the second excitation manifold (labelled by $\mathrm{n}$, see Fig.~\ref{fig1}b). Hence, when this nonlinear system is driven by a laser at the fundamental cavity resonance (evidenced by the arrow in Fig.~\ref{fig1}b), the first photon couples easily to the incoming radiation while the second one cannot enter the cavity, as there is no available state. Thus, a single-photon blockade is realized in close analogy to Kerr-type nonlinear systems \cite{imamoglu99,ferretti2012}. However, this condition holds as long as the excitation power remains small, since at larger pumping rate higher excitation manifolds become occupied, and more photons eigenstates start appearing at the empty cavity resonance, effectively destroying the blockade effect. This qualitative understanding will shortly be confirmed by rigorous numerical results.}

In order to fully describe the physical scheme proposed in Fig.~\ref{fig1}, we assume to pump the fundamental cavity mode with an input laser of frequency $\omega_L$
at a rate $\Omega$, which is described by
\begin{equation}\label{total_ham}
\hat{H}= \hat{H}_s + \hbar\Omega(t) e^{-i \omega_L t}\, \hat{a}^\dag + \hbar\Omega^{\ast}(t) e^{i \omega_L t} \, \hat{a} \, .
\end{equation}
The system dynamics can be rotated with respect to the laser frequency, $\omega_L$, by  applying the operator
$\hat{R}(t)=\mathrm{exp}\{i (\omega_L t \,\hat{a}^{\dagger}\hat{a} + 2 \omega_L t \,\hat{b}^{\dagger}\hat{b}) \}$,
which gives an effective hamiltonian $\hat{H}_{\mathrm{eff}}=\hat{R}\hat{H}\hat{R}^\dag -i\hbar \hat{R}({\mathrm{d}\hat{R}^\dag}/{\mathrm{d}t}) $, i.e.
\begin{eqnarray}\label{rotated_ham}
\hat{H}_{\mathrm{eff}}&=&
 \hbar\Delta_a \hat{a}^{\dagger}\hat{a} +  \hbar\Delta_b \hat{b}^{\dagger}\hat{b}
+ \hbar g_{\mathrm{nl}} [\hat{b}(\hat{a}^\dag)^2+\hat{b}^\dag \hat{a}^2 ]   \nonumber \\
&+&  \hbar\Omega(t) \hat{a}^\dag + \hbar\Omega^{\ast}(t) \hat{a} \, .
\end{eqnarray}
In the last expression, $\Delta_a=\omega_a-\omega_L $, $\Delta_b=\omega_b-2\omega_L $ are the detunings of the fundamental
and second-harmonic modes from the driving laser frequency and its second-harmonic, respectively.
Losses can be described within a master equation treatment for the rotated density matrix, $\tilde{\rho}=\hat{R} \rho \hat{R}^\dag$,
in Markov approximation
\begin{equation}\label{master_eq}
\frac{ \mathrm{d} \tilde{\rho}}{\mathrm{d}t}=\frac{i}{\hbar}[\tilde{\rho},\hat{H}_{\mathrm{eff}}] + \mathcal{L}(\kappa_{a,b})+\mathcal{L}_{d}(\gamma_{a,b}) \, ,
\end{equation}
where $\mathcal{L} = \sum_{i=a,b} \kappa_i [\hat{A}_i \tilde{\rho} \hat{A}_i^{\dagger} - \hat{A}_i^{\dagger}\hat{A}_i \tilde{\rho}/2-\tilde{\rho}\hat{A}_i^{\dagger} \hat{A}_i /2]$ and
$\mathcal{L}_d = \sum_{i=a,b} \gamma_{i} [\hat{A}_i^{\dagger} \hat{A}_i  \tilde{\rho} \hat{A}_i^{\dagger} \hat{A}_i - (\hat{A}_i^{\dagger} \hat{A}_i )^2 \tilde{\rho}/2-\tilde{\rho} (\hat{A}_i^{\dagger} \hat{A}_i )^2 /2]$ are the Lindblad operators corresponding to the intrinsic ($\kappa_{a,b}$) and pure dephasing ($\gamma_{a,b}$) loss rates of the two modes,
respectively. In the absence of pure dephasing, the quality factors of the two resonant modes is defined as $Q_a = \omega_a / \kappa_a$ and $Q_b = \omega_b / \kappa_b$
(i.e. $Q_b = 2 \omega_a / \kappa_b$ in this case), respectively. {We notice that we did not take into account any nonlinear loss mechanism, such as two-photon absorption. This is justified when assuming weak pumping, under which conditions these effects can safely be neglected \cite{ferretti2012}.}

The figure of merit quantifying the single-photon sensitivity for this quantum nonlinear system is
the time-ordered second-order autocorrelation function, defined as \cite{loudon_book}
\begin{equation}\label{g2_function}
G_i^{(2)}(t,t') = \langle \hat{A}_i^{\dag} (t) \hat{A}_{i}^{\dag} (t') \hat{A}_{i} (t') \hat{A}_{i} (t) \rangle  \,
\end{equation}
with $t'-t= \tau > 0$ and $i=a,b$, or its normalized version
\begin{equation}\label{g2_normal}
g_i^{(2)}(t,t') = \frac{G_i^{(2)}(t,t')}  {\langle \hat{A}_i^{\dag} (t) \hat{A}_{i} (t)\rangle  \langle \hat{A}_{i}^{\dag} (t') \hat{A}_{i} (t') \rangle} \, .
\end{equation}
In the following, we numerically calculate the latter quantities by assuming realistic parameters for state-of-art nanostructured resonators.
Mutual correlations between the fundamental and second-harmonic photons can also be investigated through the
cross-correlation function
\begin{equation}\label{g2_cross}
g_{ab}^{(2)}(t,t') = \frac{\langle \hat{a}^{\dag} (t) \hat{b}^{\dag} (t') \hat{b} (t') \hat{a}(t) \rangle}
{\langle \hat{a}^{\dag} (t) \hat{a}(t)\rangle  \langle \hat{b}^{\dag} (t') \hat{b} (t') \rangle} \, .
\end{equation}
Under continuous wave (cw) excitation, $\Omega(t)=\Omega_0$, we will mainly refer to the steady
state zero-time delay second order correlation, i.e.
$g_{i}^{(2)}(0) = \langle\hat{A}_i^{\dag2}\hat{A}_i^{2}\rangle/ \langle\hat{A}_i^{\dag}\hat{A}_i\rangle^2 =
\mathrm{Tr}\{\hat{A}_i^{\dag2}\hat{A}_i^{2} \tilde{\rho}_{ss} \} / n_i^2$,
where $n_i=\mathrm{Tr}\{\hat{A}_i^{\dag}\hat{A}_i \tilde{\rho}_{ss} \}$, and $\tilde{\rho}_{ss}$ is the steady state solution corresponding to
$ {\mathrm{d} \tilde{\rho}}/{\mathrm{d}t} = 0$.
Under pulsed excitation, e.g. for a train of Gaussian pulses $\Omega(t)=\Omega_0 \mathrm{exp}\{-(t-nT_0)^2/\Delta T^2\}$ ($n=0,\pm 1,...$),
where $T_0$ and $\Delta T$ are the pulse separation and width, respectively,
Eq. \ref{g2_function} needs to be evaluated numerically through \cite{mete2004,ciuti06prb}
\begin{equation}\label{g2_pulsed}
G_i^{(2)}(t,t') = \mathrm{Tr}\{\hat{A}_i \mathcal{U}_{t,t'}[\hat{A}_i \tilde{\rho}(t)\hat{A}_i^{\dag} ] \hat{A}_i^{\dag} \}  \, ,
\end{equation}
where $\mathcal{U}_{t,t'}$ indicates the evolution from $t$ to $t'$ with Eq. \ref{master_eq}, for an initial condition given by the operator
$\hat{A}_i \tilde{\rho}(t)\hat{A}_i^{\dag}$.

\section{ Nonlinear coupling }\label{sec:nonlinear}

For an order of magnitude estimate of the coupling rate between fundamental and second-harmonic photons,
we can simplify the expression in Eq. \ref{shift_chi2} as
\begin{equation}\label{shift_chi2_simp}
\hbar g_{\mathrm{nl}} =  \varepsilon_0 \left(\frac{\hbar\omega_a }{\varepsilon_0 \varepsilon_r }\right)^{3/2}  \frac{\bar{\chi}^{(2)}}{\sqrt{V_{\mathrm{mode}}}}
\, ,
\end{equation}
where we have assumed: $\omega_b = 2 \omega_a$, $D=2$ (conservative assumption), ${\alpha}_a(\mathbf{r})={\alpha}_b(\mathbf{r})={\alpha} $,
i.e. perfect spatial alignment and overlap between the two harmonic modes (optimistic assumption).
Within our formalism, an effective mode volume for the scalar field profile is defined in Eq. \ref{shift_chi2_simp} as
$V_{\mathrm{mode}}^{-1/2} = \int [\alpha(\mathbf{r})]^3 \mathrm{d}^3\mathbf{r} $. {We point out that such definition for the mode volume does not coincide with the usual one employed in cavity QED \cite{andreani99prb}. This is a consequence of the effective mode volume being defined according to the specific nonlinear process under examination, and different definitions might lead to slightly different quantitative estimates \cite{notomi_review}.}
An analytic estimation can be given for a normalized function describing the field confinement in a nanostructure cavity. With reference to photonic crystal resonators at near infrared wavelengths \cite{notomi_review}, we assume $\alpha(\mathbf{r})=(2/\pi\sigma_{x}\sigma_{y}d)^{1/2} \exp(-x^{2}/2\sigma_{x}^{2}-y^{2}/2\sigma_{y}^{2})cos(\pi/d)z$, which satisfies the
normalization condition above, and for which one analytically gets $ V^{1/2}_{\mathrm{mode}} = \sqrt{4 \pi \sigma_x \sigma_y d /3 } $ as a function of the relevant confinement lengths along the three spatial directions. In Eq. \ref{shift_chi2} we have made the further approximation of considering averaged values for the dielectric permittivity and the nonlinear susceptibility of the resonator, which we will take equal to the bulk values of the related material henceforth.

In photonic crystal resonators made of high-index non-centrosymmetric materials, values can be as small as \cite{robinson05prl,arakawa2011}: $\sigma_x =\sigma_y = d= \lambda/ 2 \sqrt{\varepsilon_r}$, $\varepsilon_r = 12$. With these values, and assuming $\lambda = 1.5$ $\mu$m, the mode volume estimated from the expression above is $V_{\mathrm{mode}} \simeq 0.04$ $\mu$m$^3$. For a typical second-order susceptibility $\bar{\chi}^{(2)} \simeq 2 \times 10^{-10}$ m/V, which holds for III-V semiconductor compounds (see e.g. \onlinecite{bergfeld2003prl}), we get a realistic order of magnitude estimate for the coupling constant in Eq. \ref{shift_chi2} as $\hbar g_{\mathrm{nl}} \simeq 2$ $\mu$eV, which is a remarkably large value for a passive nonlinear material. To solve the master equation for this model system, we have to assume a realistic value for cavity modes loss rates. Neglecting the effects of pure dephasing for the moment being, we take $Q_a = 2 Q_b =7 \times 10^5$, and hence  $\hbar \kappa_a = \hbar \kappa_b /4 \simeq 1$ $\mu$eV at $1.5$ $\mu$m. Such values have been experimentally demonstrated for fundamental photonic crystal cavity modes at $1.55$ $\mu$m in state-of-art nanostructured cavities made of III-V materials \cite{derossi2008}, and we have assumed that second-harmonic modes will have a Q-factor that is at least a factor of 2 smaller than the fundamental mode~\cite{andreani06prb}.

{In summary, it should be noted that the scheme proposed here relies on the simultaneous realization of three main conditions for the nanostructured cavity: high Q-factor of fundamental mode, small mode volume, and double resonance. As discussed above, these conditions have been shown (either experimentally or theoretically) to occur separately in different types of photonic crystal cavities, but not simultaneously at time of writing. However, the fast pace of advancement of cavity design and nanostructuring capabilities is likely to produce the required conditions within the same device in the near future, and holds promise for the present proposal to stimulate further efforts in this direction. }

\section{CW excitation}\label{sec:cw}

\subsection{Dependence on cavity parameters}

\begin{figure}[t]
\begin{center}
\includegraphics[width=0.42\textwidth]{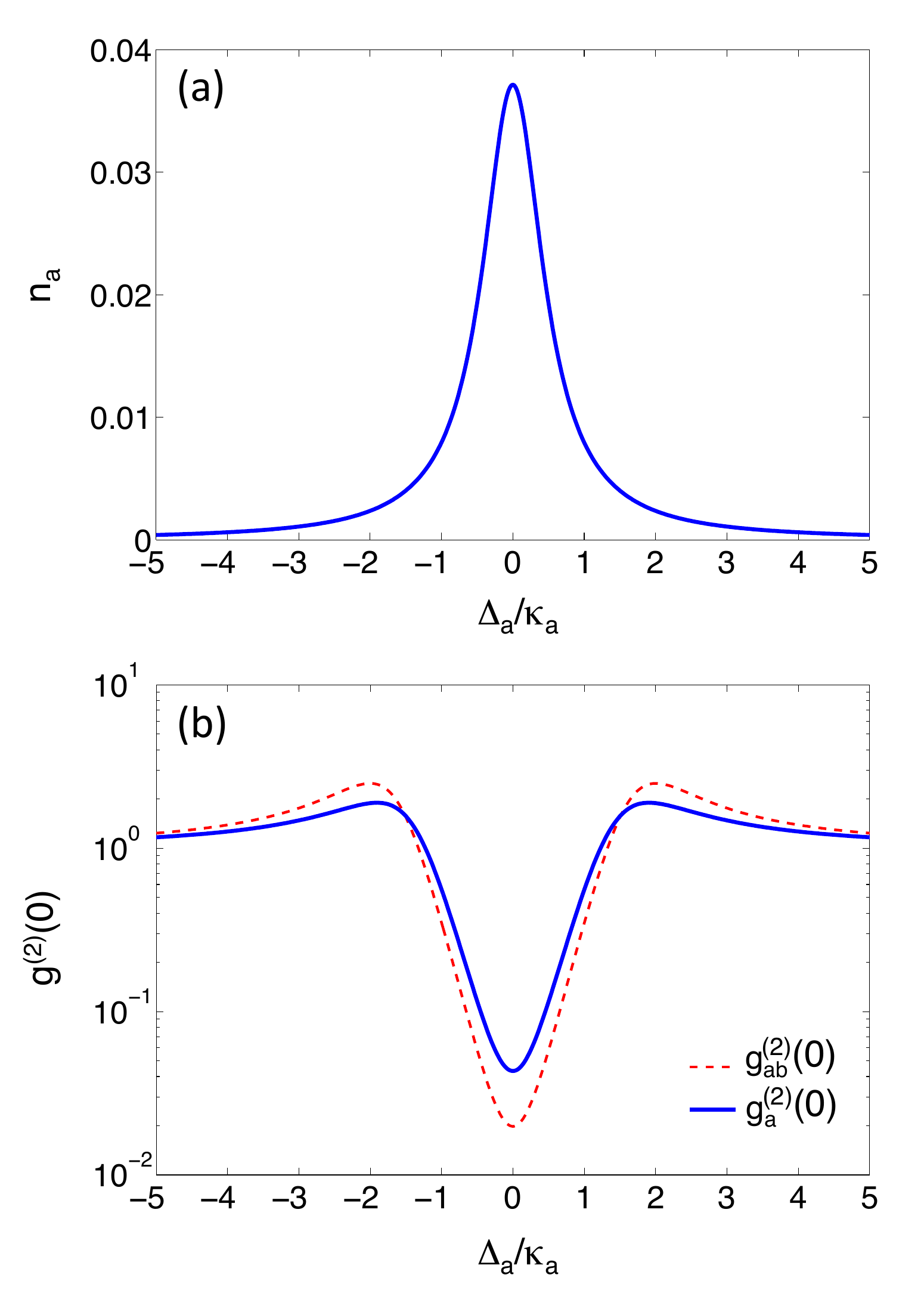}
\caption{(Color online) Average photon population in the fundamental cavity mode at frequency $\omega_a$ as a function of laser frequency detuning (a), and corresponding auto- and cross-correlation functions between modes a and b at zero time delay, respectively (b). Parameters: $g_{\mathrm{nl}}/ \kappa_a =2$, $\kappa_b = 4 \kappa_a$, $\Omega_0 / \kappa_a = 0.1$.} \label{fig2}
\end{center}
\end{figure}

With the parameters estimated above, we have solved the master equation for the steady state density matrix and calculated the second-order correlation functions (both auto- and cross-correlations between fundamental and second-harmonic modes) at zero-time delay as a function of the pump/cavity mode detuning, with a cw pump at constant rate $\Omega_0$. The results are shown in Fig.~\ref{fig2}. The average number of photons in the cavity at frequency $\omega_a$ follows an almost Lorentzian resonance, as shown in  Fig.~\ref{fig2}a. Correspondingly, a strong antibunching signal can be detected in the autocorrelation, Fig.~\ref{fig2}b, for a laser frequency exactly tuned with the fundamental cavity mode (i.e., $g_a^{(2)} (0) = 0.043 $ in the minimum of the antibunching dip). Under such conditions, the system acts as an efficient single-photon blockade device for the laser photons. In the same plot we also show the cross-correlation between first and second harmonic photons, displaying a strong anticorrelated signal. The latter indicates a suppressed probability for the simultaneous presence of single-photons at $\omega_a$ and $2\omega_a$, respectively. As a consequence, single-photon emission can occur alternatively at fundamental or second-harmonic frequency, although the latter process is far less efficient than the first (owing to the reduced population at $2\omega_a$).
The spectral broadening of the antibunching dip in Fig.~\ref{fig2}b evidently reflects the low-power resonance broadening in Fig.~\ref{fig2}a. {The small bunching that is visible for $g_a^{(2)} (0)$ around $\Delta_a / \kappa_a \sim \pm \sqrt{2}$ derives from the incoming laser resonantly driving the second excitation manifold, which occurs at $\omega_L \simeq \omega_a \pm \sqrt{2} g_{\mathrm{nl}} /2$ (see Fig.~\ref{fig1}b).}

\begin{figure}[t]
\begin{center}
\includegraphics[width=0.45\textwidth]{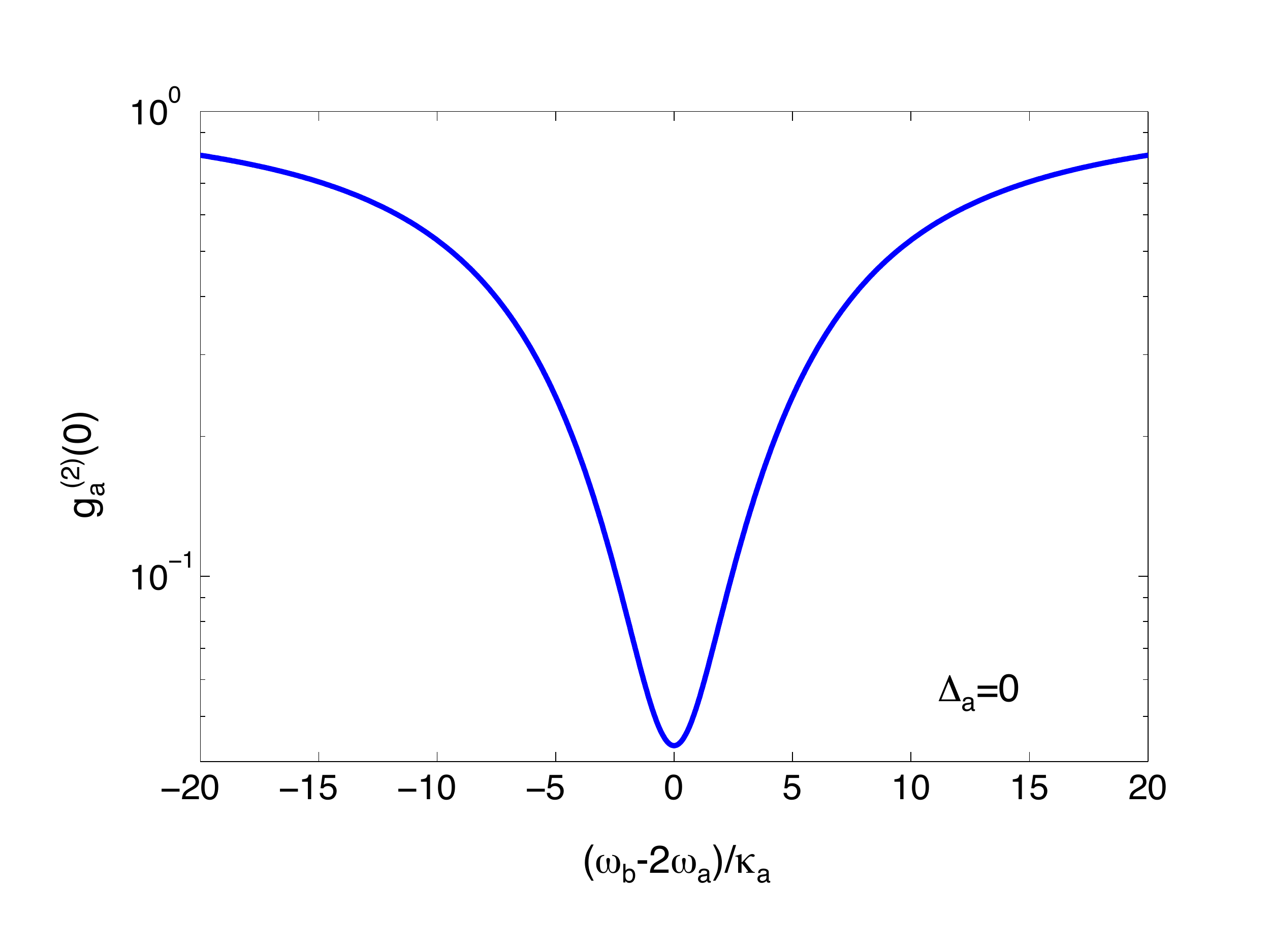}
\caption{(Color online) Auto-correlation function at zero time delay for the fundamental mode with resonant laser, as a function of second-order mode frequency. Parameters of the calculation: $g_{\mathrm{nl}}/ \kappa_a =2$, $\kappa_b = 4 \kappa_a$, $\Omega_0 / \kappa_a = 0.1$.} \label{fig3new}
\end{center}
\end{figure}

{ In Fig.~\ref{fig3new} we show the tolerance of the antibunching dip for resonant excitation, i.e. $\omega_L = \omega_a$, on the second-order mode deviation from the ideal second-harmonic condition, i.e. $g_a^{(2)} (0)$ as a function of $\Delta_b - 2 \Delta_a = \omega_b - 2\omega_a$, for fixed $\omega_a$. As it can be seen from the numerical results, the single-photon blockade of the input laser is preserved up to several linewidths (e.g., at least $\pm 5 \kappa_a$) deviation of second-order mode from the second-harmonic frequency. This is promising in view of possible realizations of this proposal, where exact matching of the frequency $\omega_b$ with the condition $2\omega_a$ is likely to be dependent on fabrication tolerances of realistic devices (see also discussion in the previous section). }

\begin{figure}[t]
\begin{center}
\vspace{0.1cm}
\includegraphics[width=0.42\textwidth]{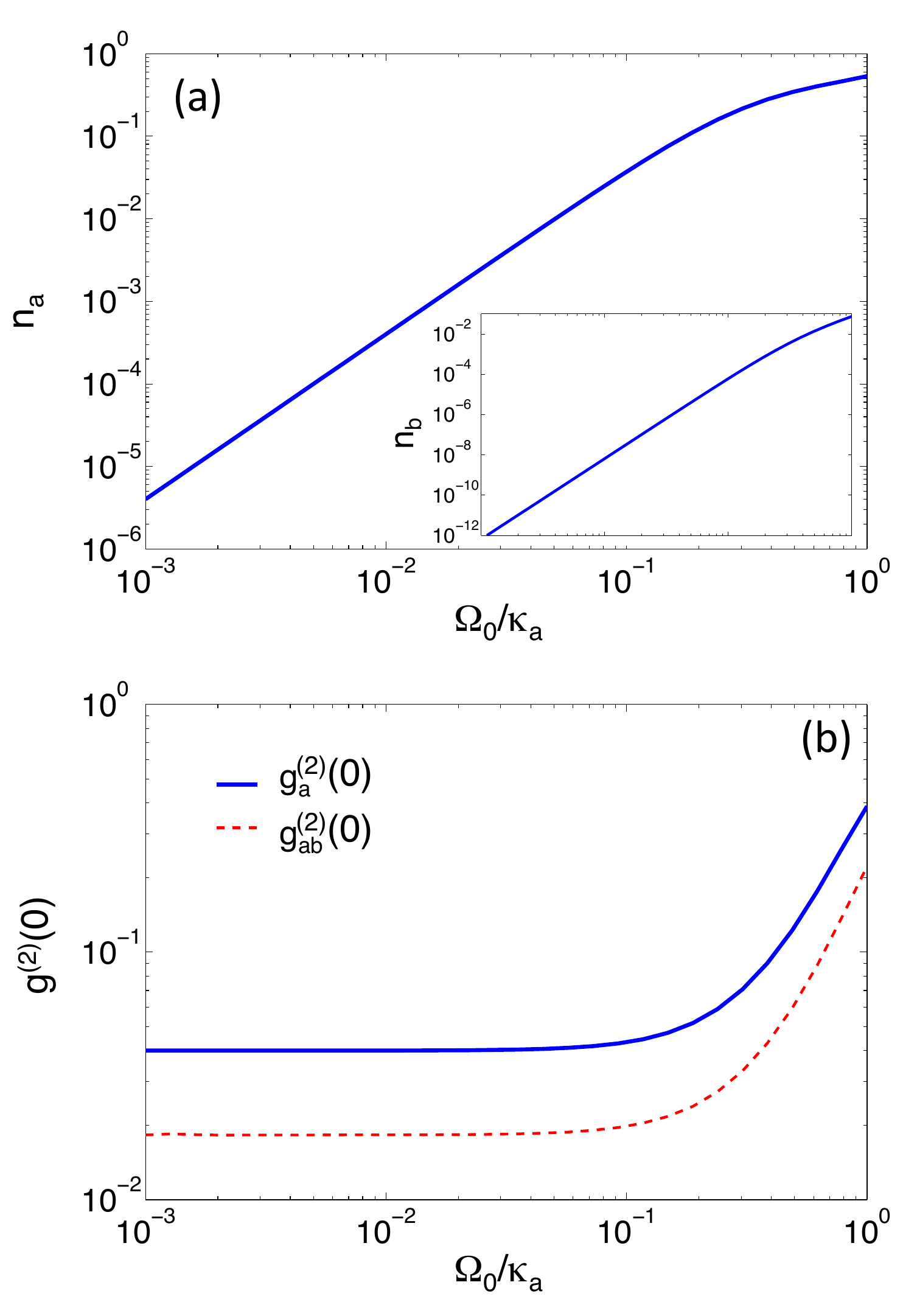}
\caption{(Color online) Average photon population in the fundamental cavity mode as a function of pumping rate (a), and corresponding auto- and cross-correlation functions at zero time delay between modes a and b (b). The inset of panel (a) shows the average population of the second-harmonic mode. Parameters: $g_{\mathrm{nl}}/ \kappa_a =2$, $\kappa_b = 4 \kappa_a$, $\Delta_a=\Delta_b=0$.} \label{fig3}
\end{center}
\end{figure}

So far, we have assumed a pumping rate $\Omega_0 \ll \kappa_a$ in the calculations to ensure remaining in the weakly nonlinear regime. Hence, from the results in Fig.~\ref{fig2} we can estimate the maximum single-photon emission rate at the driving laser frequency ($\omega_L = \omega_a$) as $n_a \kappa_a \simeq 1$ GHz, which is comparable to single-photon sources based, e.g., on single quantum dots \cite{sps_review} with a typical lifetime on the order of nanoseconds.
{ Dependence of the average number of photons in the fundamental mode on the driving strength is shown in Fig.~\ref{fig3}a. At small pumping rate, the second-harmonic mode population is negligibly small as compared to the fundamental mode, as shown in the inset. This a posteriori justifies neglecting terms proportional to $n_b$ in Eq.~\ref{system_ham}.  
On increasing the pump power, the antibunching dip is suppressed, as explicitly shown in Fig.~\ref{fig3}b. Such behavior can be attributed to the fact that with increasing driving the number of excitations in the system increases. This leads to the appearance of eigenstates at the empty cavity resonance, as previously discussed on the basis of the schematic level scheme reported in Fig.~\ref{fig1}b. Such behavior is similar to the single-photon blockade mechanism in a cQED system described by the Jaynes-Cummings model, which behaves classically (i.e., does not emit antibunched radiation) under strong driving \cite{AM_PRA_R}.
From the results shown in this figure, it is evident that the device works efficiently in the single-photon blockade regime only at low pump powers of about $\Omega_0 / \kappa_a \leq 0.1$. }

\begin{figure}[t]
\begin{center}
\includegraphics[width=0.48\textwidth]{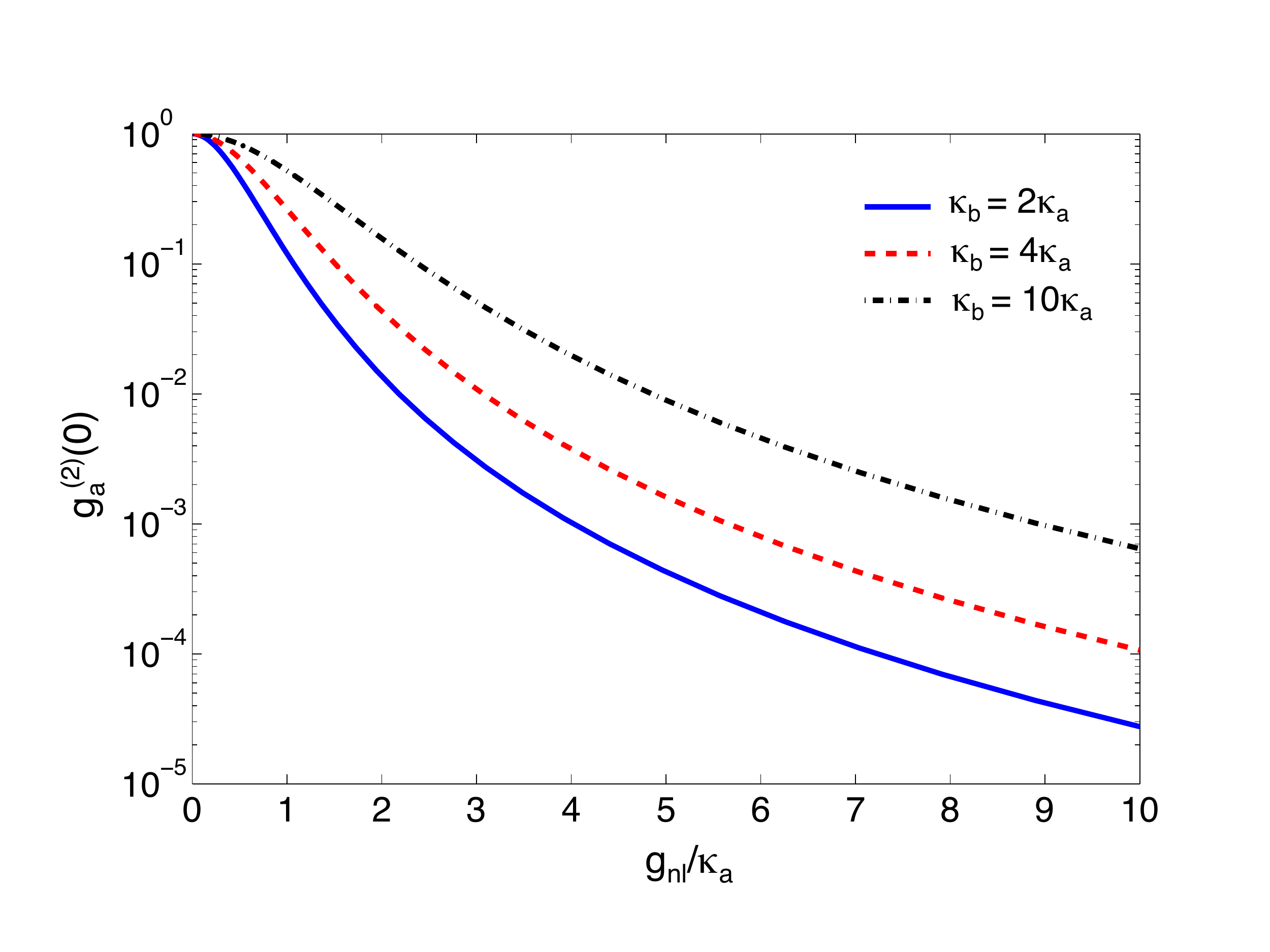}
\caption{(Color online) Dependence of the antibunching dip for the fundamental mode on the effective two-photon nonlinearity, for different second-harmonic mode loss rates. Parameters used for these calculations are: $\Omega_0 / \kappa_a = 0.1$, $\Delta_a=\Delta_b=0$.} \label{fig4}
\end{center}
\end{figure}

Clearly, the efficiency of such a passive device as a cw single-photon source is strongly dependent on the effective value of the nonlinear coupling rate between fundamental and second-harmonic modes, $g_{\mathrm{nl}}$. Even if we have used a realistic estimate as given in Sec.~\ref{sec:nonlinear}, we show in Fig.~\ref{fig4} how the antibunching explicitly depends on the ratio $g_{\mathrm{nl}}/ \kappa_a$. As it can be seen, $g_a^{(2)}\to 0$ quite rapidly for $g_{\mathrm{nl}}/ \kappa_a >1$, similarly to the Kerr-type nonlinearity~\cite{ferretti2012}. From the results in Fig.~\ref{fig4} it can also be inferred that single-photon blockade is preserved to a good extent even when decreasing the second-harmonic Q-factor, especially if $g_{\mathrm{nl}}/ \kappa_a$ is kept large enough: even if the second-harmonic mode has a five-fold lower Q-factor than the fundamental mode (i.e., $\kappa_b = 10 \kappa_a$), $g_a^{(2)} (0) < 0.1$ already for $g_{\mathrm{nl}} / \kappa_a \simeq 3$.
From a practical point of view, there are different possibilities to increase the ratio $g_{\mathrm{nl}} / \kappa_a$ in realistic devices. First of all, we have based our analysis on a Q-factor for the fundamental mode $Q_a < 10^6$, while larger Q-factors have already been shown in the literature at near infrared wavelengths in the 1.5 $\mu$m band (see \onlinecite{notomi_review} for a recent review). Second, the actual values of $g_{\mathrm{nl}}$ depend on a number of details of the specific cavity design or material choice, as it is immediately evident from Eq.~\ref{shift_chi2}. In any case, the results shown here are a good indication that interesting quantum photonics applications might be in order for passive devices based on second-order nonlinear materials, provided a doubly resonant nanocavity can be designed and realized.

\subsection{Effect of pure dephasing}

We now analyze the effects of a detrimental source of decoherence in the system, such as pure dephasing of the cavity modes. Such an effect is the analog of phonon-assisted scattering or spectral diffusion for solid state quantum emitters, which produces a line broadening of the resonance lineshapes, whose effects on single-photon sources based on cavity QED systems have been fully characterized (see, e.g., Ref. \onlinecite{auffeves2010}). In the present context, any source of pure dephasing on the cavity photons (i.e. acting on coherences rather than populations) is potentially going to affect the degree of antibunching, and hence the effective usefulness for single-photon generation. From a physical point of view, three main sources of pure dephasing can be identified in passive devices: (i) thermal instabilities (depending on the experimental conditions), (ii) coupling to mechanical vibrations of the nano cavity (important, e.g., for suspended resonators such as photonic crystal membranes), and (iii) stability of the pumping laser wavelength. Overall, these effects can be described by the Lindblad term $\mathcal{L}_{d}(\gamma_{a,b})$ in Eq.~\ref{master_eq}, in which we assume the same pure dephasing rate $\gamma_d = \gamma_{a,b}$ for fundamental and second-harmonic modes, respectively.
Even if in such a general context it is difficult to attribute specific pure dephasing rates, the latter depending on the system under investigation, nevertheless we give here an indication on how the antibunching dip in the second-order autocorrelation essentially depends on the magnitude of this source of decoherence.

\begin{figure}[t]
\begin{center}
\vspace{0.5cm}
\includegraphics[width=0.42\textwidth]{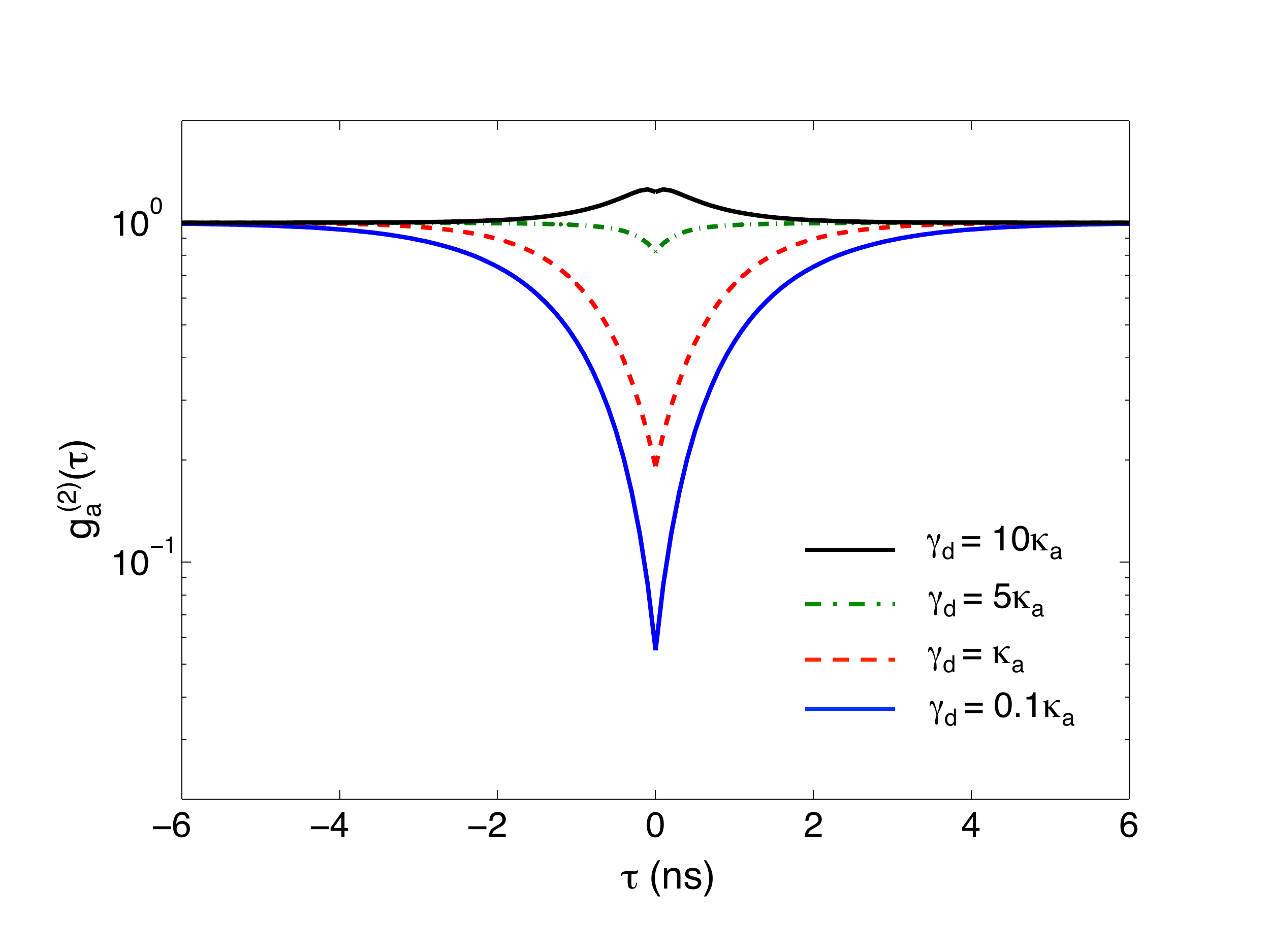}
\caption{(Color online) Steady state second-order autocorrelation of the fundamental mode as a function of time delay ($\tau=t'-t$, with $t \to \infty$), for different values of the pure dephasing rate. Parameters: $\hbar \kappa_a = 1$ $\mu$eV, $g_{\mathrm{nl}}/ \kappa_a =2$, $\kappa_b = 4 \kappa_a$, $\Omega_0 / \kappa_a = 0.1$, $\Delta_a=\Delta_b=0$.} \label{fig5}
\end{center}
\end{figure}

In Fig.~\ref{fig5} we show the time-dependent second order autocorrelation in steady state for the fundamental cavity mode, under resonant cw excitation at rate $\Omega_0 = 0.1 \kappa_a$ and for a pure dephasing rate $\gamma_d$ (given in units of the population decay rate at $\omega_a$). For these calculations, we have assumed a realistic cavity decay rate for the fundamental mode, $\kappa_a / 2\pi \simeq 1.5$ GHz (corresponding to $Q_a \simeq 7 \times 10^5$ at 1.5 $\mu$m wavelength). The behavior of $g_a^{(2)} (\tau)$ closely mimics the typical one from a cw coherently driven two-level system, with a recovery time determined by the internal photon lifetime (in this case set by $2\pi/ \kappa_a \simeq 0.66$ ns).  As it can be seen from the figure, the antibunching dip is only weakly affected for values of pure dephasing rates on the order of $\gamma_d \leq 0.1 \kappa_a$, where $g_a^{(2)} (0)\simeq 0.055$ has to be compared to the value of 0.043 of Fig.~\ref{fig3} for the same pumping rate and $\gamma_d = 0$. On increasing the pure dephasing rate, the antibunching dip is progressively reduced, and finally the system behaves as a bunched light source similar to a thermal radiator. However, it should be emphasized that we indeed expect typical pure dephasing rates to be $\gamma_d < 0.1 \kappa_a$ in realistic devices, even at room temperature. Overall, such results confirm a substantial robustness of the present model to decoherence of the photon field inside the resonator.

\section{Pulsed excitation}\label{sec:pulsed}

Single-photon sources on demand require the emission of single-photon pulses at deterministic times. The most straightforward way of obtaining such a source employs a sequence of pulses with spectral pulse-width smaller than the energy scale set by the single-photon nonlinearity in the system (i.e., $\hbar g_{\mathrm{nl}}$ in our case), and temporal pulse separation much larger than the cavity photon lifetime. Single-photon sources on demand have been characterized by a master equation treatment similar to the one employed here for either cavity QED schemes \cite{mete2004} or Kerr-type nonlinear cavities \cite{ciuti06prb}, respectively. Recently, highly efficient such sources have been experimentally demonstrated by using semiconductor quantum dots emitting in the $1$ $\mu$m wavelength range at cryogenic temperature \cite{mete2013nnano}.

\begin{figure}[t]
\begin{center}
\includegraphics[width=0.48\textwidth]{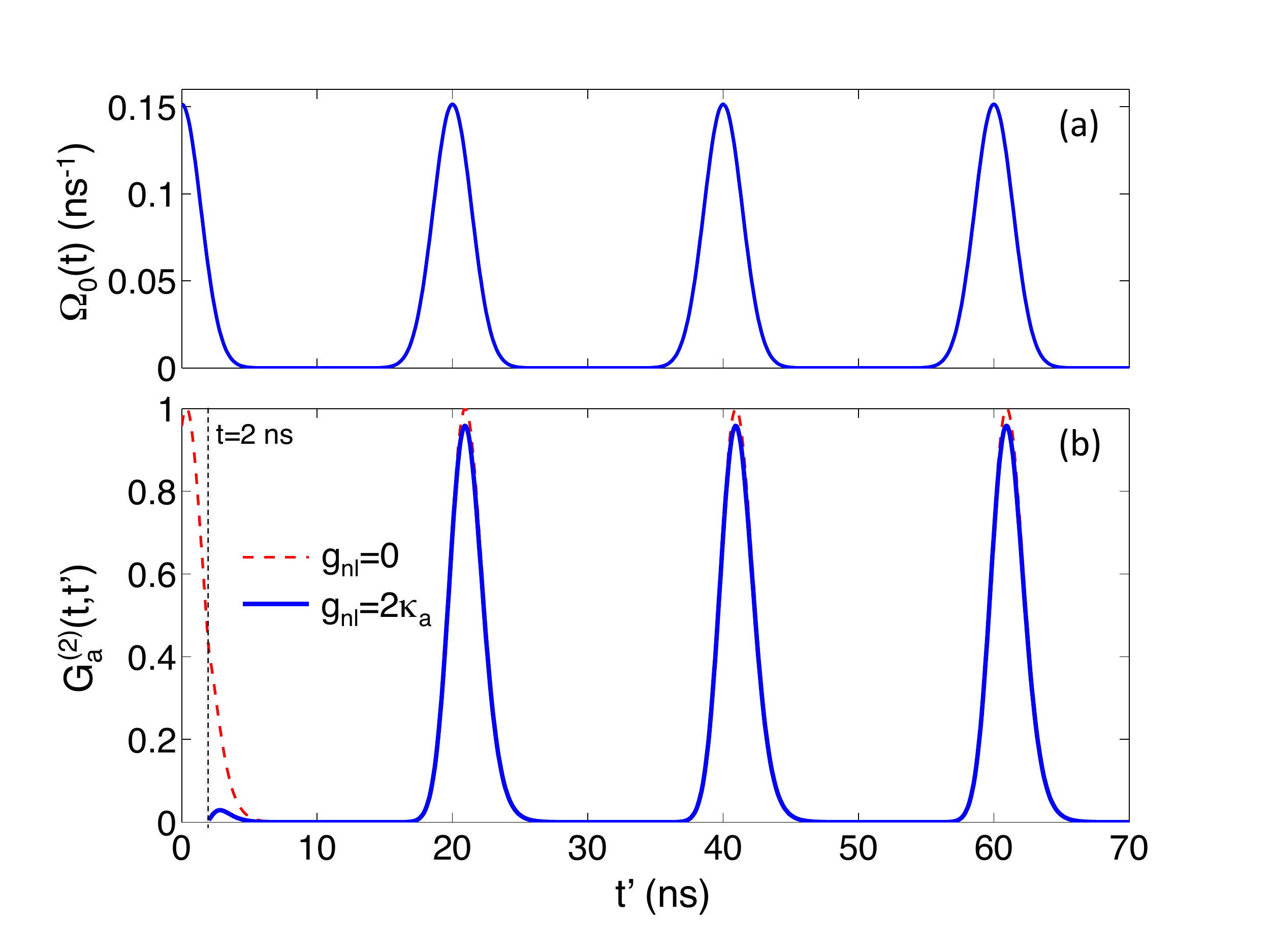}
\caption{ (Color online) Sequence of gaussian pulses with $\Delta T = 2$ ns width and $T_0=20$ ns separation (a), and corresponding unnormalized second-order correlation function, calculated from Eq.~\ref{g2_pulsed}, for $t=2$ ns in the cases with (full line) and without (dashed) nonlinear interaction (b).  Parameters: $\hbar \kappa_a = 1$ $\mu$eV, $\kappa_b = 4 \kappa_a$, $\Omega_0 / \kappa_a = 0.1$, $\Delta_a=\Delta_b=0$. } \label{fig6}
\end{center}
\end{figure}

Here we show that the doubly resonant cavity scheme with second-order nonlinearity  is straightforwardly suited for experiments under pulsed excitation, thereby promising this system as a potential passive single-photon source on demand for operation at arbitrary temperature and wavelengths in the mid-infrared.
The behavior of $g_a^{(2)} (0)$ under cw excitation is a necessary condition for the device to be operated also in pulsed excitation. We explicitly show in Fig.~\ref{fig6} that by properly choosing the pulse duration and repetition, according the the system nonlinearity and lifetime, the time dependent second-order correlation shows a pronounced antibunching behavior for a train of gaussian pulses. We assume parameters as in the previous section, with a fundamental cavity photon lifetime on the order of a nanosecond. The sequence of gaussian pulses is shown in Fig.~\ref{fig6}a, with a maximum coherent pumping rate $\hbar \Omega_0 = 0.1$ $\mu$eV. Correspondingly, we show in Fig.~\ref{fig6}b the numerically calculated $G_a^{(2)} (t,t')$, for $t=2$ ns and for positive time delay $\tau > 0$. We notice that we do not show the normalized $g_a^{(2)} (t,t')$, since the cavity population is depleted after each pulse. However, the information on the intensity correlation is already contained in the un-normalized quantity, $G_a^{(2)} (t,t')$. For a second-order nonlinear coupling on the order of $\hbar g_{\mathrm{nl}}=2 $ $\mu$eV, as assumed in the previous Section, we correctly get a suppression of the zero time-delay peak in the two-time correlation signal, as expected for a pulsed single-photon source \cite{mete2004,ciuti06prb}. We checked that by setting the nonlinear coupling $g_{\mathrm{nl}}=0$, no suppression occurs. The area of the suppressed peak reflects the value of the antibunching dip as obtained under cw excitation.

\section{Conclusions}

In conclusion, we propose the use of nanostructured non-centrosymmetric materials with strong $\chi^{(2)}$ contribution to their nonlinear susceptibility for applications in quantum photonics. The proposal relies on the ability to design and fabricate doubly resonant photonic nanocavities with small confinement volume and large quality factor for both fundamental and second-harmonic resonances. Although interesting designs have recently been proposed for doubly resonant ring resonators~\cite{johnson2012opex} and might be worth applying the present concepts to these systems as well, mode volumes are still too large in such devices; thus, it seems that most promising systems to realize our proposal could be photonic crystal cavities etched in thin nonlinear semiconductor slabs with high index contrast \cite{notomi_review,rivoire2011opex,andreani06prb}, {where all of the optimal conditions to realize our proposed model are likely to be simultaneously achieved in the near future.}
By assuming state of the art parameters for III-V semiconductor cavities, we have shown that strong antibunching can be achieved at the output of the device for resonant coherent driving with a laser frequency tuned to the fundamental mode resonance, both with continuous wave and pulsed excitations, respectively. Such antibunching is a signature of single-photon emission, for which we have checked the robustness against the main sources of decoherence.

The results reported in the present paper show the usefulness of passive nanophotonic devices as single-photon sources on demand, with potential impact in telecom-range and room-temperature quantum photonic circuits~\cite{qp_review} where the lack of suitable integrated quantum emitters might hinder large-scale diffusion in the long term. We thus believe it is worth realizing doubly resonant passive nanocavities, as a promising alternative to ongoing efforts in developing integrated single-photon sources by using different types of quantum emitters, such as single molecules, defects in solids, or quantum dots.

\begin{acknowledgments}
Useful and stimulating discussions with S. Buckley, I. Carusotto, M. Liscidini, V. Savona, J. Vu\v{c}kovi\'{c}, and F. Wang  are gratefully acknowledged.
\end{acknowledgments}


\end{document}